\documentclass[preprint,12pt]{elsarticle}
\usepackage{amsthm}
\usepackage{amsmath}
\usepackage{amssymb}
\usepackage{amstext}
\usepackage{bm}
\usepackage{graphicx}
\usepackage{epstopdf}
\usepackage{subcaption}
\usepackage{hyperref}

 \newcommand{\bl}{\big<}
  \newcommand{\bg}{\big>}
 
\biboptions{sort&compress}

\journal{---}

\begin{document}

\begin{frontmatter}

\title{ON THE ACCURACY OF THE NON-CLASSICAL TRANSPORT EQUATION IN 1-D RANDOM PERIODIC MEDIA}

\author[ufrgs]{Richard Vasques\corref{cor1}}
\author[rwth]{Kai Krycki\fnref{presad}}
\cortext[cor1]{
richard.vasques@fulbrightmail.org
}
\fntext[presad]{krycki@mathcces.rwth-aachen.de}
\address[ufrgs]{PROMEC - School of Engineering\\
UFRGS - Federal University of Rio Grande do Sul, Porto Alegre, Brazil}
\address[rwth]{Department of Mathematics,
Center for Computational Engineering Science,\\
RWTH Aachen University, Aachen, Germany}

\begin{abstract}
We present a first numerical investigation of the accuracy of the recently proposed {\em non-classical transport equation}. This equation contains an extra independent variable (the path-length $s$), and models particle transport taking place in random media in which a particle's distance-to-collision is {\em not} exponentially distributed. To solve the non-classical equation, one needs to know the $s$-dependent ensemble-averaged total cross section $\Sigma_t(s)$, or its corresponding path-length distribution function $p(s)$. We consider a 1-D spatially periodic system consisting of alternating solid and void layers, randomly placed in the infinite line. In this preliminary work, we assume transport in rod geometry: particles can move only in the directions $\mu=\pm 1$. We obtain an analytical expression for $p(s)$, and use this result to compute the corresponding $\Sigma_t(s)$. Then, we proceed to solve the non-classical equation for different test problems. To assess the accuracy of these solutions, we produce ``benchmark" results obtained by (i) generating a large number of  physical realizations of the system, (ii) numerically solving the transport equation in each realization, and (iii) ensemble-averaging the solutions over all physical realizations. We show that the results obtained with the non-classical equation accurately model the
ensemble-averaged scalar flux in this 1-D random system, generally outperforming the widely-used atomic mix model. We conclude by discussing plans to extend the present work to slab geometry, as well as to more general random mixtures.

\end{abstract}

\begin{keyword}
random media \sep non-classical transport \sep atomic mix \sep pebble bed


\end{keyword}

\end{frontmatter}

\section{Introduction}\label{sec1}
\setcounter{section}{1}
\setcounter{equation}{0} 

 In the classical theory of linear particle transport, the probability density function for a particle's distance-to-collision is given by an exponential:
$p(s) = \Sigma_te^{-\Sigma_ts}$,
where the total cross section $\Sigma_t$ is independent of the path-length $s$ (the distance traveled by the particle since its
previous interaction) and of the direction of flight $\bm\Omega$.
  
However, in certain random media in which the locations of the scattering centers are spatially correlated, the
particle flux will experience a non-exponential attenuation law.
Examples include neutron transport in Pebble Bed Reactors (in which a non-exponential $p(s)$ arises due to the pebble arrangement within the core) and photon transport in atmospheric clouds (in which the locations of the water droplets in the cloud seem to be correlated in ways that measurably affect the radiative transfer within the cloud).
 
An approach to this type of ``non-classical'' transport problem was recently introduced \cite{lar_07,larvas_11}, with the assumption that the positions of the scattering
centers are correlated but independent of direction $\bm \Omega$;
existence and
uniqueness of solutions are
rigorously discussed in \cite{fra_10}. The non-classical theory was extended \cite{vaslar_14a} to include angular-dependent path-length distributions. Furthermore, a similar kinetic equation with path-length as an independent variable has been rigorously derived for the periodic Lorentz gas in a series of papers by Golse et al.\ (cf.\ \cite{gol_12}) as well as Marklof and Str\"ombergsson (cf.\  \cite{mar_11}).

In the case of isotropic scattering, the non-classical linear Boltzmann equation with angular-dependent path-length distributions and isotropic source is writen as
\begin{align}\label{eq1}
\frac{\partial\psi}{\partial s}(\bm x,\bm\Omega,s) + \bm\Omega\cdot\bm\nabla \psi(\bm x,\bm\Omega,s) &+ \Sigma_t(\bm\Omega,s)\psi(\bm x,\bm\Omega,s) 
\\&= \frac{\delta(s)}{4\pi}\left[ c\int_{4\pi}\int_0^\infty \Sigma_t(\bm\Omega',s')\psi(\bm x,\bm\Omega',s')ds' d\Omega' + Q(\bm x) \right]\,, \nonumber
\end{align}
where $\psi$ represents the angular flux, $c$ is the scattering ratio, and $\Sigma_t(\bm\Omega,s)$ is the angular-dependent ensemble-averaged total cross section, defined as
\begin{equation}\label{eq2}
\Sigma_t(\bm\Omega,s)ds =  \begin{array}{l}
\text{ the probability (ensemble-averaged over all physical realizations) that a}\\
\text{ particle, scattered or born at any point $\bm x$ and traveling in the direction $\bm\Omega$,}\\
\text{ will experience a collision between $\bm x + s\bm\Omega$ and $\bm x + (s+ds)\bm\Omega$.}\\
\end{array}
 \end{equation}
The underlying path-length distribution and the above non-classical cross section are related \cite{vaslar_14a} by
\begin{align}\label{eq3}
	p(\bm\Omega,s) = \Sigma_t(\bm\Omega,s)\exp\left( -\int_0^s\Sigma_t(\bm\Omega,s')ds'\right).
\end{align}
Numerical results have been provided for the asymptotic diffusion limit of this non-classical theory
\cite{larvas_11,vaslar_09,vas_13,vaslar_14b}, and for moment models of the non-classical equation in the diffusive regime \cite{kry_13}. However, to the best of our knowledge, no results have been presented to the non-classical \textit{transport} equation. This is because one must know $\Sigma_t(\bm\Omega,s)$, or $\Sigma_t(s)$ in the case of angular-independent path lengths, in order to solve Eq.\ (\ref{eq1}). 

In this paper, we present a first numerical investigation of the accuracy of the non-classical transport equation. We consider a 1-D random periodic system: a spatially periodic system consisting of alternating solid and void layers, randomly placed in the infinite line $-\infty<x<\infty$. This means that we only know which material is present at any given point $x$ in a probabilistic sense. In this case, Eq.\ (\ref{eq1}) can be written as
\begin{align}\label{eq4}
\frac{\partial\psi}{\partial s}(x,\mu,s) + \mu\frac{\partial \psi}{\partial x}(x,\mu,s) &+ \Sigma_t(\mu,s)\psi(x,\mu,s) 
\\& = \frac{\delta(s)}{2}\left[ c\int_{-1}^1\int_0^\infty \Sigma_t(\mu',s')\psi(x,\mu',s')ds' d\mu' + Q(x) \right]\,. \nonumber
\end{align}
For this system, we can obtain an analytical expression for $p(\mu,s)$, the distribution function for a particle's distance-to-collision in the direction $\mu$. Then, using the identity \cite{vaslar_14a}
\begin{align}\label{eq5}
\Sigma_t(\mu,s)=\frac{p(\mu,s)}{1-\int_0^sp(\mu,s')ds'},
\end{align}
we can obtain a solution for Eq.\ (\ref{eq4}).

In this preliminary work, we assume transport in {\em rod geometry}, in which particles can only move in the directions $\mu=\pm 1$. We consider 6 different test problems, and provide numerical results for the non-classical equation. To analyze the accuracy of these results, we compare them against ``benchmark" numerical results, obtained by ensemble-averaging the solutions of the transport equation over a large number of physical realizations of the random system. Finally, we compare the performance of the non-classical model against the widely-known atomic mix model.

The remainder of this paper is organized as follows. In Section \ref{sec2} we present the steps to analytically obtain the path-length distribution function $p(s)$ in the 1-D random periodic system. In Section \ref{sec3} we (i) define the test problems; (ii) discuss the models used to solve the test problems; and (iii) present and analyze the numerical results. We present our conclusions in Section \ref{sec4}.    

\section{Path-length Distribution Function in the 1-D System}\label{sec2}
\setcounter{section}{2}
\setcounter{equation}{0} 

\begin{figure}[h]
  \centering
  \includegraphics[width=\textwidth]{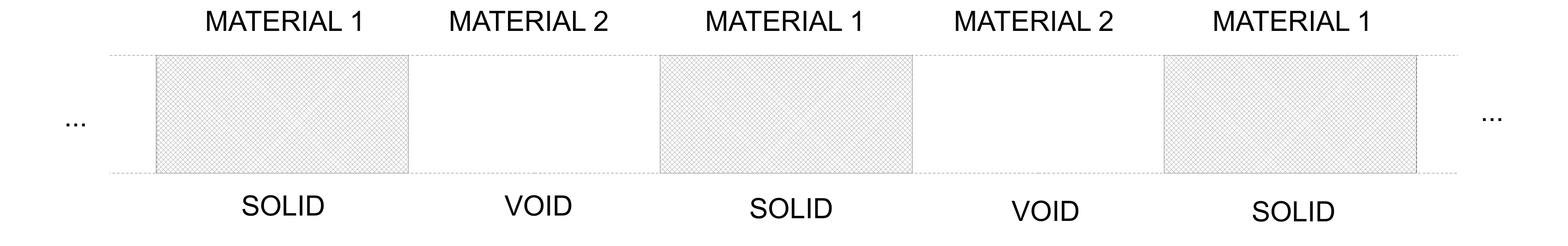}
  \caption{A sketch of the periodic medium}
  \label{fig1}
\end{figure}

 We consider a 1-D physical system consisting of alternating layers of two distinct materials, periodically arranged. We assume material 1 to be the one in which particles can be born and collide, while  material 2 is assumed to be void. In this preliminary work, the thickness $\ell$ of the layers of both materials is assumed to be the same;
a sketch of the periodic system is given in Figure \ref{fig1}. 

This periodic system is {\em randomly placed} in the infinite line $-\infty < x < \infty$, such that the probability $P_i$ of finding material $i$ in a given point $x$ is $1/2$. Therefore, the cross sections and source are stochastic functions of space:
\begin{align}\label{eq6}
\Sigma_t(x) =  \left\{
\begin{array}{cl}
\Sigma_{t1}, & \text{if $x$ is in material 1}\\
0, &\text{if $x$ is in material 2}\\
\end{array}
\right.
\, , \,\,
Q(x) =  \left\{
\begin{array}{cl}
Q_1, & \text{if $x$ is in material 1}\\
0, &\text{if $x$ is in material 2}\\
\end{array}
\right.
.
\end{align}

Given a physical realization of this system, let us consider a particle that is born (or scatters) at a point $x$ in a layer of material 1 with direction of flight $\mu\neq 0$. We take $x_0$ to be the horizontal distance between $x$ (the point in which the collision or birth event took place) and the next intersection between layers in the direction $\mu$, and define:
\begin{subequations}\label{eq7}
\begin{align}
a(x_0,\mu,s)ds &= \begin{array}{l}
\text{ the probability that a particle released at a horizontal distance $x_0$ of the}\\
\text{  next intersection and traveling in the direction $\mu$ will experience a}\\
\text{ collision between $s$ and $s+ds$;}\\
\end{array}
\\
q(x_0,\mu,s) &= \begin{array}{l}
\text{ the probability that a particle released at a horizontal distance $x_0$ of the}\\
\text{  next intersection and traveling in the direction $\mu$ will travel a distance}\\
\text{ $s$ without colliding.}\\
\end{array}
\end{align}
\end{subequations}
Therefore, for $\mu\neq 0$ we can write
\begin{subequations}\label{eq8}
\begin{align}
a(x_0,\mu,s) &=  \left\{
\begin{array}{ll}
\Sigma_{t1}, & \text{if } 0\leq s\leq x_0/|\mu|\\
0, &\text{if } x_0/|\mu|< s\leq x_0/|\mu|+\ell/|\mu|\\
\Sigma_{t1}, &\text{if } x_0/|\mu|+\ell/|\mu|< s\leq x_0/|\mu|+2\ell/|\mu|\\
\,\,\,\vdots & \\
\end{array}
\right.
\,\,;\\\nonumber\\
q(x_0,\mu,s) &=  \left\{
\begin{array}{ll}
e^{-\Sigma_{t1}s}, & \text{if } 0\leq s\leq x_0/|\mu|\\
e^{-\Sigma_{t1}x_0/|\mu|}, &\text{if } x_0/|\mu|< s\leq x_0/|\mu|+\ell/|\mu|\\
e^{-\Sigma_{t1}(s-\ell/|\mu|)}, &\text{if } x_0/|\mu|+\ell/|\mu|< s\leq x_0/|\mu|+2\ell/|\mu|\\
e^{-\Sigma_{t1}(x_0/|\mu|+\ell/|\mu|)}, &\text{if } x_0/|\mu|+2\ell/|\mu|< s\leq  x_0/|\mu|+3\ell/|\mu|\\
e^{-\Sigma_{t1}(s-2\ell/|\mu|)}, &\text{if } x_0/|\mu|+ 3\ell/|\mu|< s\leq x_0/|\mu|+4\ell/|\mu|\\
e^{-\Sigma_{t1}(x_0/|\mu|+2\ell/|\mu|)}, &\text{if } x_0/|\mu|+4\ell/|\mu|< s\leq  x_0/|\mu|+5\ell/|\mu|\\
\,\,\,\vdots & \\
\end{array}
\right.
\,\,.
\end{align}
\end{subequations}
For simplicity, in this paper we will only consider transport in {\em rod geometry}; that is, particles are only allowed to travel in the horizontal directions $\mu = \pm 1$. This eliminates the angular dependence of the probability functions defined in Eqs.\ (\ref{eq8}), since they must be even functions of $\mu$. In this case, the general forms of Eqs.\ (\ref{eq8}) (for $n=0,1,2, ...$)  are given by
\begin{subequations}\label{eq9}
\begin{align}
a(x_0,s) &=  \left\{
\begin{array}{ll}
\Sigma_{t1}, & \text{if } 0\leq s\leq x_0\\
0, &\text{if } x_0+2n\ell< s\leq  x_0+(2n+1)\ell\\
\Sigma_{t1}, &\text{if } x_0+(2n+1)\ell< s\leq x_0+2(n+1)\ell\\
\end{array}
\right.
\,\,;\\\nonumber\\
q(x_0,s) &=  \left\{
\begin{array}{ll}
e^{-\Sigma_{t1}s}, & \text{if } 0\leq s\leq x_0\\
e^{-\Sigma_{t1}(x_0+n\ell)}, &\text{if } x_0+2n\ell< s\leq  x_0+(2n+1)\ell\\
e^{-\Sigma_{t1}[s-(n+1)\ell]}, &\text{if } x_0+(2n+1)\ell< s\leq x_0+2(n+1)\ell\\
\end{array}
\right.
\,\,.
\end{align}
\end{subequations}
In order to obtain the ensemble-averaged probabilities over all physical realizations, we operate on Eqs.\ (\ref{eq9}) by $\frac{1}{\ell}\int_0^\ell (\cdot) dx_0$, which gives
\begin{subequations}\label{eq10}
\begin{align}
a(s) &=  \left\{
\begin{array}{ll}
\frac{\Sigma_{t1}}{\ell}\left[(2n+1)\ell-s\right], & \text{if } 2n\ell\leq s\leq (2n+1)\ell\\
-\frac{\Sigma_{t1}}{\ell}\left[(2n+1)\ell-s\right], &\text{if } (2n+1)\ell\leq s\leq 2(n+1)\ell\\
\end{array}
\right.
\,\,;\\\nonumber\\
q(s)&=  \left\{
\begin{array}{ll}
\frac{1}{\ell}\left[\left((2n+1)\ell-s-\frac{1}{\Sigma_{t1}}\right)e^{-\Sigma_{t1}(s-n\ell)} +\frac{e^{-\Sigma_{t1}n\ell}}{\Sigma_{t1}}\right], &\hspace{-1,5cm} \text{if } 2n\ell\leq s\leq (2n+1)\ell\\
-\frac{1}{\ell}\left[\left((2n+1)\ell-s-\frac{1}{\Sigma_{t1}}\right)e^{-\Sigma_{t1}[s-(n+1)\ell]} +\frac{e^{-\Sigma_{t1}(n+1)\ell}}{\Sigma_{t1}}\right], &\\
&\hspace{-2.5cm}\text{if } (2n+1)\ell\leq s\leq 2(n+1)\ell\\
\end{array}
\right.
\end{align}
\end{subequations}
Finally, keeping in mind that
\begin{align}\label{eq11}
p(s)ds &= \begin{array}{l}
\text{probability that a particle will experience its first collision}\\
\text{while traveling a distance between $s$ and $s+ds$}\\
\end{array}
\\&= \,\,\,a(s)ds\times q(s),\nonumber
\end{align}
the path-length distribution function in this 1-D random periodic system is given by
\begin{align}\label{eq12}
p(s)&=  \left\{
\begin{array}{ll}
\frac{\Sigma_{t1}}{\ell^2}\left[(2n+1)\ell-s\right]
\left[\left((2n+1)\ell-s-\frac{1}{\Sigma_{t1}}\right)e^{-\Sigma_{t1}(s-n\ell)} +\frac{e^{-\Sigma_{t1}n\ell}}{\Sigma_{t1}}\right], &\\
&\hspace{-4,5cm} \text{if } 2n\ell\leq s\leq (2n+1)\ell\\
&\\
\frac{\Sigma_{t1}}{\ell^2}\left[(2n+1)\ell-s\right]
\left[\left((2n+1)\ell-s-\frac{1}{\Sigma_{t1}}\right)e^{-\Sigma_{t1}[s-(n+1)\ell]} +\frac{e^{-\Sigma_{t1}(n+1)\ell}}{\Sigma_{t1}}\right], &\\
&\hspace{-4,5cm}\text{if } (2n+1)\ell\leq s\leq 2(n+1)\ell\\
\end{array} 
\right.
\end{align}
In the next section we present numerical results and investigate the accuracy of the non-classical approach to solve rod geometry transport problems in this system.

\section{Models and Numerical Results}\label{sec3}
\setcounter{section}{3}
\setcounter{equation}{0} 

The test problems presented in this paper consider rod geometry transport ($\mu = \pm1$) taking place in the {\em infinite} 1-D random periodic system described in Section \ref{sec2}. The classical transport equation is written as
\begin{align}\label{eq13}
\pm \frac{\partial \psi^{\pm}}{\partial x}(x) + \Sigma_t(x)\psi^{\pm}(x) 
= \frac{c\Sigma_t(x)}{2}\left[\psi^{\pm}(x)+\psi^{\mp}(x)\right]+ \frac{Q(x)}{2}\,,
\end{align}
where $\psi^{\pm}(x) = \psi(x,\mu=\pm 1)$. The stochastic parameters $\Sigma_t(x)$ and $Q(x)$ are given by Eqs.\ (\ref{eq6}), with $\Sigma_{t1} = 1.0$. We define the layer thickness for both materials to be $\ell = 1.0$; thus, the probability $P_i$ of finding material $i$ at any given point $x$ in a physical realization of the system is $1/2$. 

For each problem set, we present numerical results for 3 different choices of the scattering ratio $c$: 0.1, 0.5, and 0.9. We focus on investigating the ensemble-averaged scalar flux $\bl\phi\bg$ generated by particles that are born near the center of the random system. To achieve this, we analyze two sets of problems, defining
\begin{align}\label{eq14}
Q_1 = \left\{
\begin{array}{cl}
1.0, & \text{if} -L\leq x\leq L\\
0, &\text{otherwise}\\
\end{array}
\right.
, \,\,\,\,\,
\text{where}\,\,\,\,\,
L = \left\{
\begin{array}{ll}
0.5\,\,(=\ell/2), & \text{Problem Set \textbf{A}}\\
1.0\,\,(=\ell), &\text{Problem Set \textbf{B}}\\
\end{array}
\right.
.
\end{align}
Specifically, this means that $\psi^\pm(x)\rightarrow 0$ as $|x|\rightarrow\infty$. In order to simulate the infinite system described here,
we take the domain of the $x$-variable to be large enough to ensure a sufficient decay of the solution. All numerical calculations in this paper were performed on a spatial domain $[-20, 20]$, with vacuum boundary conditions imposed at $x=\pm 20$.

\subsection{Benchmark Model}

The random quality of the system arises from randomly placing the periodic arrangement in the infinite line $-\infty<x<\infty$. For instance, to obtain a given physical realization, one can simply choose a segment of two full layers (one of each material) and randomly place the coordinate $x=0$ in this segment. However, the number of different realizations that can be computed is limited by the choice of spatial discretization. 

In this work, we have used the diamond spatial differencing scheme with mesh interval given by $\triangle x=2^{-7}$. For each test problem,
we have simulated the corresponding $256$ realizations of the periodic system, obtained by shifting the $x$-coordinates by $\triangle x$ each time. We have solved Eq.\ (\ref{eq13}) for the angular flux in each of the physical realizations, 
obtaining the scalar flux $\phi(x) = \psi^+(x)+\psi^-(x)$.
Finally, we have calculated the ensemble-averaged benchmark scalar flux $\bl\phi_B\bg(x)$ by averaging over all $256$ physical realizations. 

In all problems, differences in the numerical results for $\bl\phi_B\bg(x)$ were negligible when increasing the number of mesh intervals and the number of realizations; thus, we have concluded that these benchmark results are adequately accurate for the scope of this preliminary work.

\subsection{The Atomic Mix Model}

The {\em atomic mix model} \cite{pom_91,gol_00} consists of replacing, in the classical transport equation, the
stochastic parameters (cross sections and source) by their volume-averages. In 1-D geometry, this model is known to be accurate when the material layers are optically thin. The atomic mix equation in rod geometry for the problems discussed in this paper is given by
\begin{subequations}\label{eq15}
\begin{align}\label{eq15a}
\pm \frac{\partial \bl\psi^{\pm}\bg}{\partial x}(x) + \bl\Sigma_t\bg\bl\psi^{\pm}\bg(x) 
= \frac{c\bl\Sigma_t\bg}{2}\left[\bl\psi^{\pm}\bg(x)+\bl\psi^{\mp}\bg(x)\right]+ \frac{\bl Q\bg(x)}{2}\,,
\end{align}
where 
\begin{align}
\bl\Sigma_t\bg &= P_1\Sigma_{t1} + P_2\Sigma_{t2} = \Sigma_{t1}/2 = 0.5,\\
\bl Q\bg(x) &= P_1 Q_1 + P_2 Q_2 = Q_1/2 = \left\{
\begin{array}{cl}
0.5, & \text{if} -L\leq x\leq L\\
0, &\text{otherwise}\\
\end{array}\label{eq15c}
\right.,
\end{align}
\end{subequations}
with $L$ given in Eq.\ (\ref{eq14}).
As in the benchmark model, Eq.\ (\ref{eq15a}) is solved for the ensembled-averaged angular flux using a diamond spatial differencing scheme with mesh interval $\triangle x=2^{-7}$. The ensemble-averaged atomic mix scalar flux is given by $\bl\phi_{AM}\bg(x) = \bl\psi^+\bg(x)+\bl\psi^-\bg(x)$.

\subsection{The Non-Classical Model}
For the transport problems discussed in this work, we rewrite the non-classical Eq.\ (\ref{eq4}) in an initial value form (cf.\ \cite{lar_07}) as
\begin{subequations}\label{eq16}
\begin{align}
&\frac{\partial\psi^{\pm}}{\partial s}(x,s) \pm \frac{\partial \psi^{\pm}}{\partial x}(x,s) + \Sigma_t(s)\psi^{\pm}(x,s)  = 0, \label{eq16a}\\
& \psi^{\pm}(x,0)= \frac{c}{2} \int_0^\infty \Sigma_t(s')[\psi^{\pm}(x,s')+\psi^{\mp}(x,s')]ds' + \frac{\bl Q\bg(x)}{2}, \label{eq16b}
\end{align}
\end{subequations}
where $\psi^{\pm}(x,s) = \psi(x,\mu=\pm 1, s)$, $\bl Q\bg(x)$ is given by Eq.\ (\ref{eq15c}), and $\Sigma_t(s)$ is given by Eqs.\ (\ref{eq5}) and (\ref{eq12}).
For the numerical solution of this system, we can interpret the path-length $s$ as a pseudo-time variable. We then solve Eqs.\ (\ref{eq16}) using a finite volume method with explicit pseudo-time discretization according to \cite{hll_83}. Specifically, we adapt the scheme that was introduced in \cite{kry_13} for moment models of the non-classical transport equation. We choose a uniform grid $(x_m,s^n)$, where $x_{m+1} = x_{m} + \Delta x$ for all $m\in\mathbb{Z}$, and $s^{n+1} = s^n + \Delta s$ for all $n\in\mathbb{N}_0$. Furthermore, we define $\psi_{m}^{n,\pm}:= \psi^\pm(x_{m},s^n)$, $Q_m := \bl Q\bg (x_m)$, and $\Sigma_t^n:=\Sigma_t(s^n)$. The fully discretized system reads
\begin{subequations}\label{eq17}
\begin{align}
	& \frac{\psi^{n+1,\pm}_m-\psi^{n,\pm}_m}{\Delta s} \pm \frac{\psi^{n,\pm}_{m+1}-\psi_{m-1}^{n,\pm}}{2\Delta x} - 
			\frac{\psi^{n,\pm}_{m+1}-2\psi_m^{n,\pm}+\psi_{m-1}^{n,\pm}}{2\Delta x}   + \Sigma_t^n \psi^{n,\pm}_m = 0, \label{eq17a}\\
	&  \psi_m^{0,\pm} = \frac{c}{2}\sum\limits_{n=0}^\infty \omega_n \Sigma_t^n\left( \psi_m^{n,+} + \psi^{n,-}_m\right)  +  \frac{Q_m}{2},\label{eq17b}	
\end{align}
\end{subequations}
for some infinite quadrature rule given by the weights $\omega_n$. The second order central differences arise as a numerical diffusion term, which is typical for HLL finite volume schemes. In our calculations, we have cut off the integration at $s_{\text{max}}=80$, and have chosen the trapezoidal rule. We have used the same mesh interval $\triangle x=2^{-7}$ as for the previous models, and a CFL number $0.5$.  Because of the coupling of the initial value to the full solution in (\ref{eq16b}), this system is solved in a source-iteration manner, where we iterate between equation (\ref{eq17b}) and (\ref{eq17a}). The ensemble-averaged non-classical scalar flux is given by $\bl\phi_{NC}\bg(x) = \int_0^{80}[\psi^+(x,s)+\psi^-(x,s)]ds$.

\subsection{Numerical Results}

\begin{figure}
    \centering
    \begin{subfigure}[b]{0.495\textwidth}
        \centering
        \includegraphics[width=\textwidth]{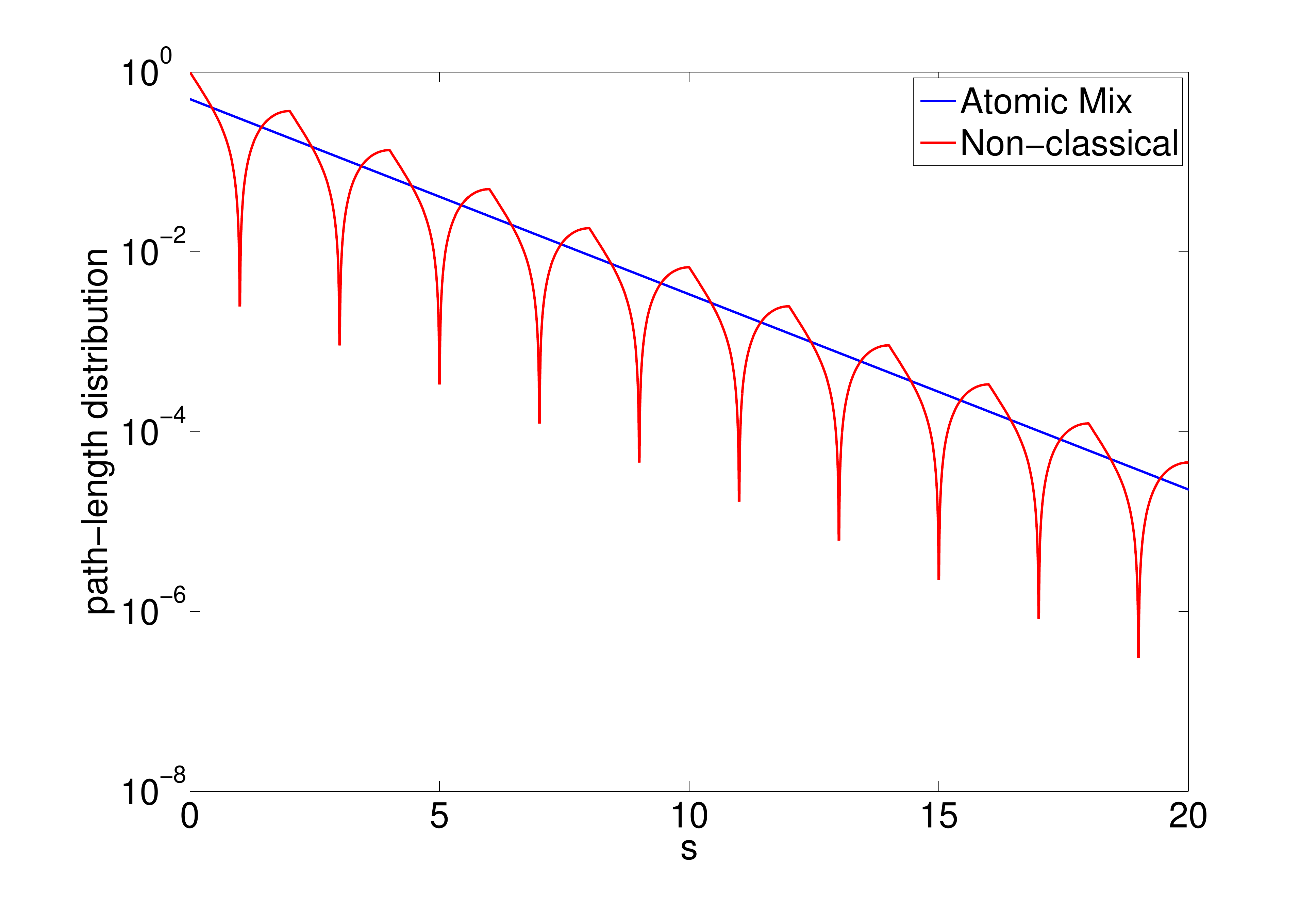}
        \caption{Path-length distributions $p(s)$ \textit{(log scale)}}
        \label{fig:PLD}
    \end{subfigure}
    \hfill
    \begin{subfigure}[b]{0.495\textwidth}
        \centering
        \includegraphics[width=\textwidth]{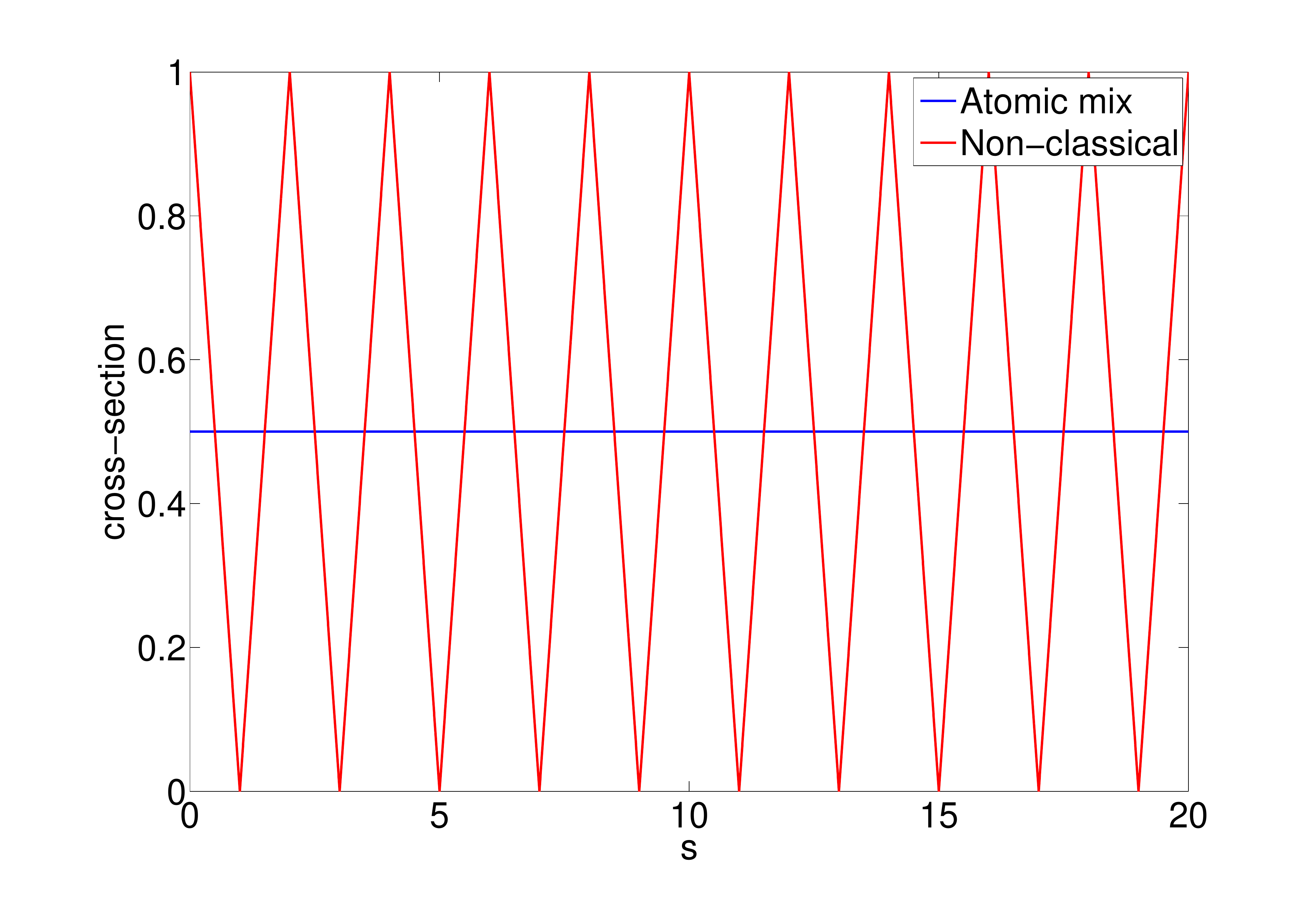}
        \caption{Total cross-setions $\bl\Sigma_t\bg$ and $\Sigma_t(s)$}
        \label{fig:Sig}
    \end{subfigure} 
    \caption{Path-length distributions and total cross-sections for the atomic mix and the non-classical models}
    \label{fig:SigPLD}
\end{figure}
The atomic mix model inherently approximates the path-length distribution function by the exponential $p(s) =  \bl\Sigma_t\bg e^{-\bl\Sigma_t\bg s}$. The non-classical model, on the other hand, uses the correct $p(s)$ that was analytically obtained in Eq.\ (\ref{eq12}). Figure \ref{fig:PLD} presents a comparison (in logarithmic scale) of the $p(s)$ used in each model for the test problems presented in this work. We remark that the local minima are supposed to have the value $0$; the deviations in the plot are caused by numerical errors.

Figure \ref{fig:Sig} shows the total cross-sections of each model as functions of the path-length $s$. Keeping in mind that $\ell=1.0$ and material 2 is a void, the ``saw-tooth" behavior of the non-classical cross-section $\Sigma_t(s)$ is consistent with the physical process, and can be easily understood:
\begin{itemize}
\item[\textbf{1.}] A particle is born or scatters in material 1. The path-length $s$ is set to 0.
\item[\textbf{2.}] At $s=0.5$, the $x$-coordinate is in material 1 with probability 1/2. Thus, $\Sigma_t(0.5) = \Sigma_{t1}/2$.
\item[\textbf{3.}] At $s=1$, the $x$-coordinate {\em must be} in material 2. Thus, $\Sigma_t(1) = \Sigma_{t2} = 0$.
\item[\textbf{4.}] At $s=1.5$, the $x$-coordinate is in material 1 with probability 1/2. Thus, $\Sigma_t(1.5) = \Sigma_{t1}/2$.
\item[\textbf{5.}] At $s=2$, the $x$-coordinate {\em must be} in material 1. Thus, $\Sigma_t(2) = \Sigma_{t1} = 1.0$.\end{itemize}
The exceptions for the scenarios we just described would be particles born exactly at {\em interface points}, which form a set of measure zero.

\begin{figure}
    \centering
    \begin{subfigure}[b]{0.495\textwidth}
        \centering
        \includegraphics[width=\textwidth]{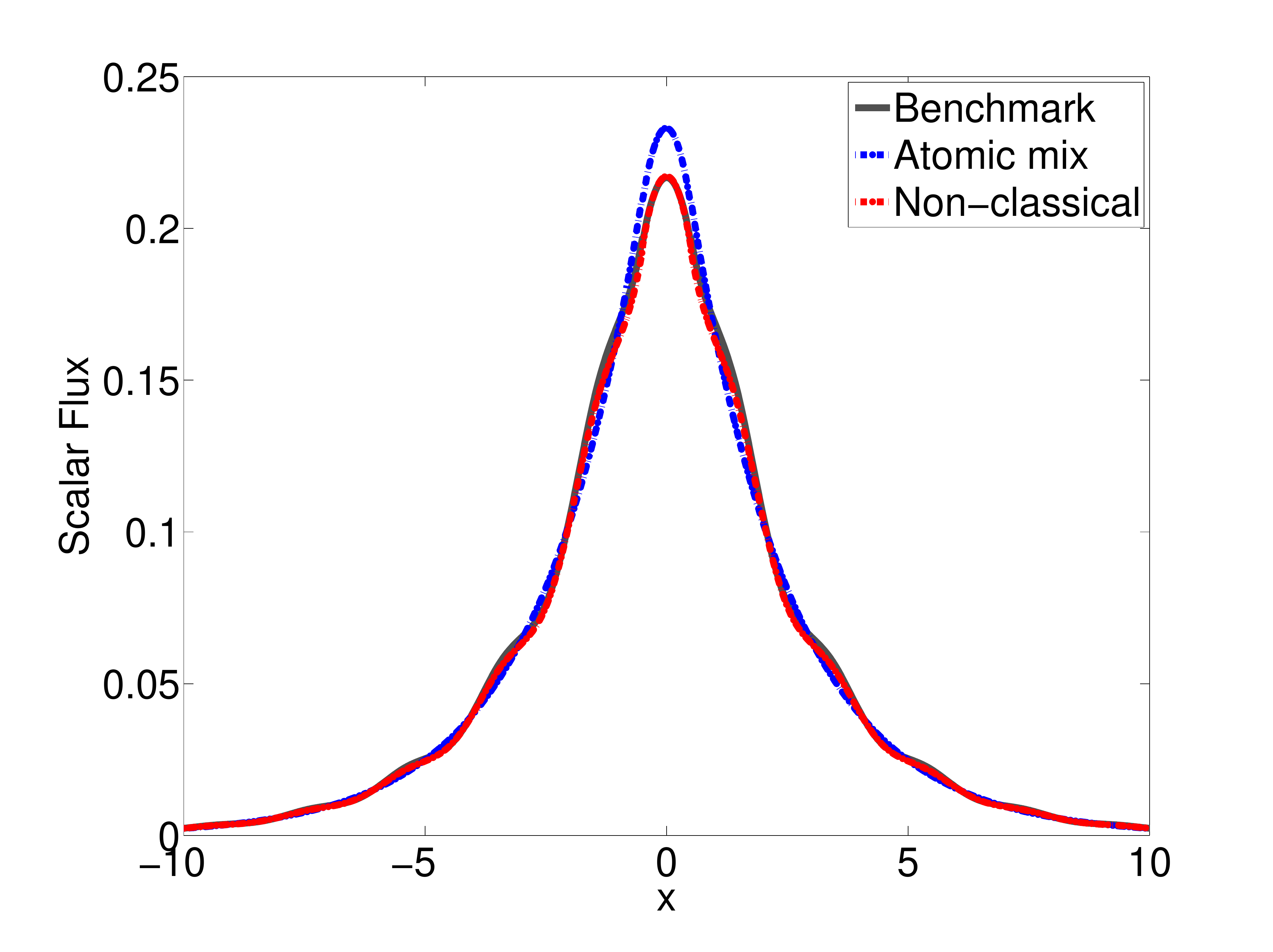}
        \caption{Problem set A, $c=0.1$}
        \label{fig:Comp09a}
    \end{subfigure}
    \hfill
    \begin{subfigure}[b]{0.495\textwidth}
        \centering
        \includegraphics[width=\textwidth]{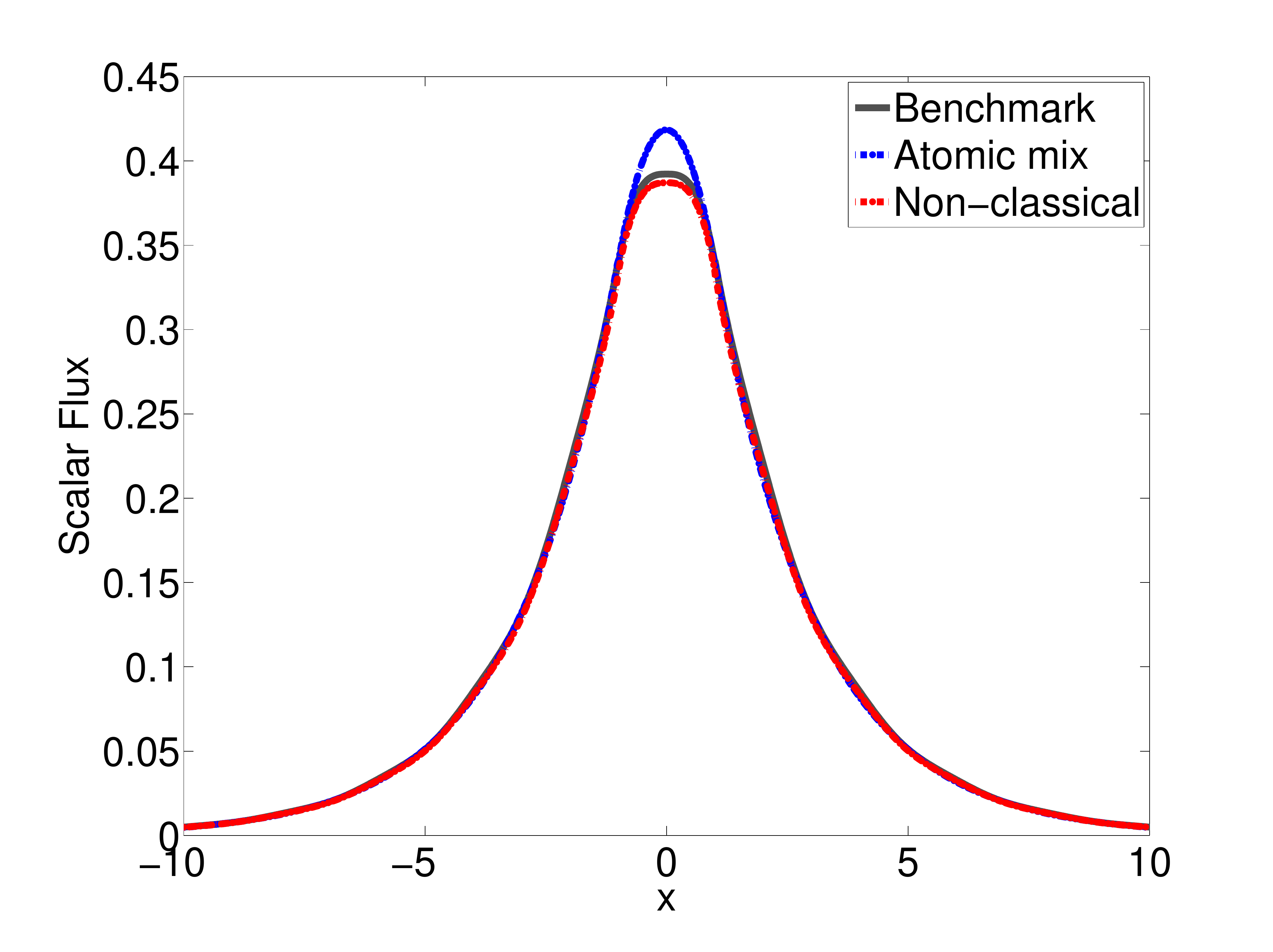}
        \caption{Problem set B, $c=0.1$}
        \label{fig:Comp09b}
    \end{subfigure}
     \\
    \begin{subfigure}[b]{0.495\textwidth}
        \centering
        \includegraphics[width=\textwidth]{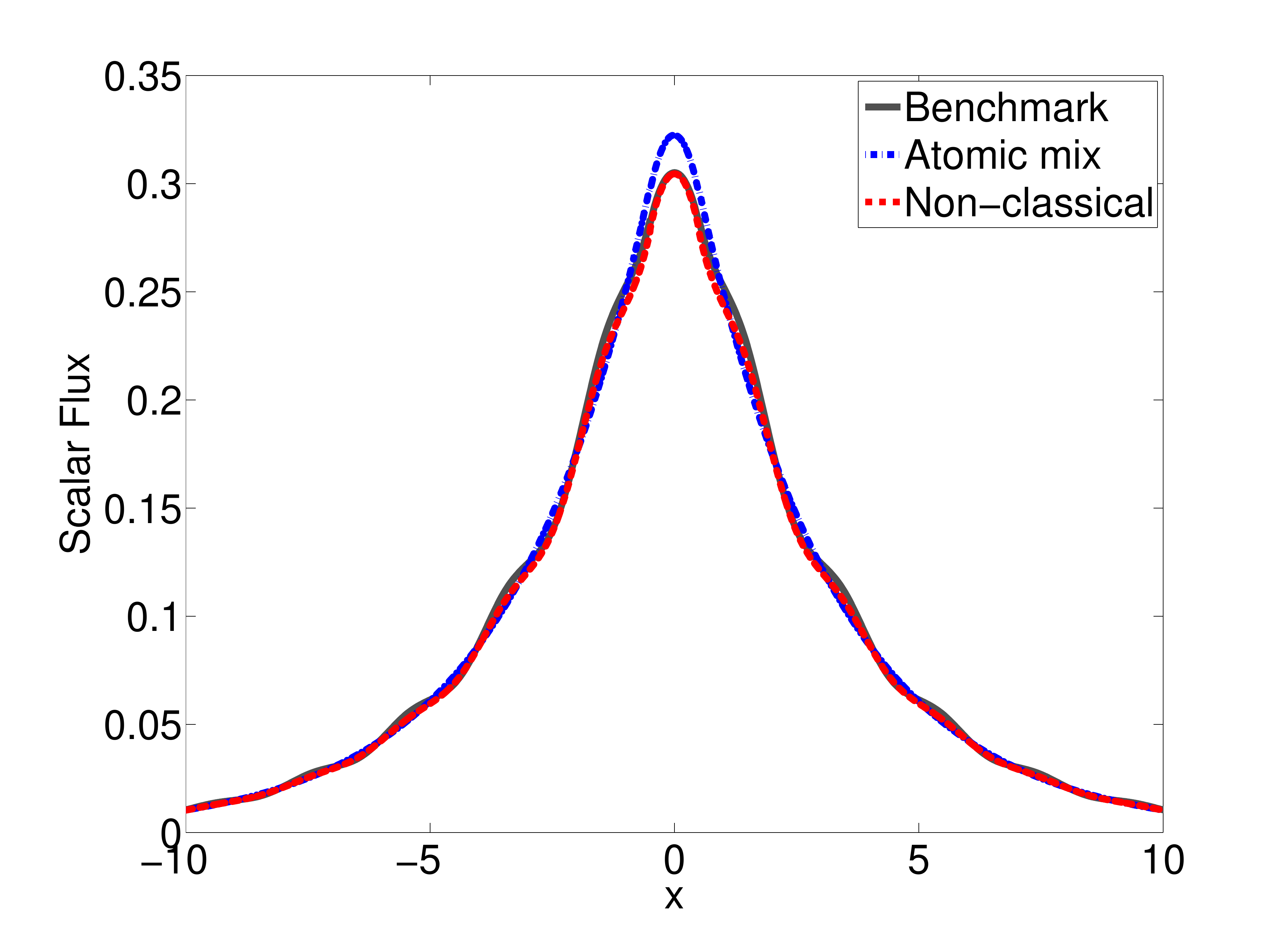}
        \caption{Problem set A, $c=0.5$}
        \label{fig:EP05a}
    \end{subfigure}
    \hfill
	 \begin{subfigure}[b]{0.495\textwidth}
        \centering
        \includegraphics[width=\textwidth]{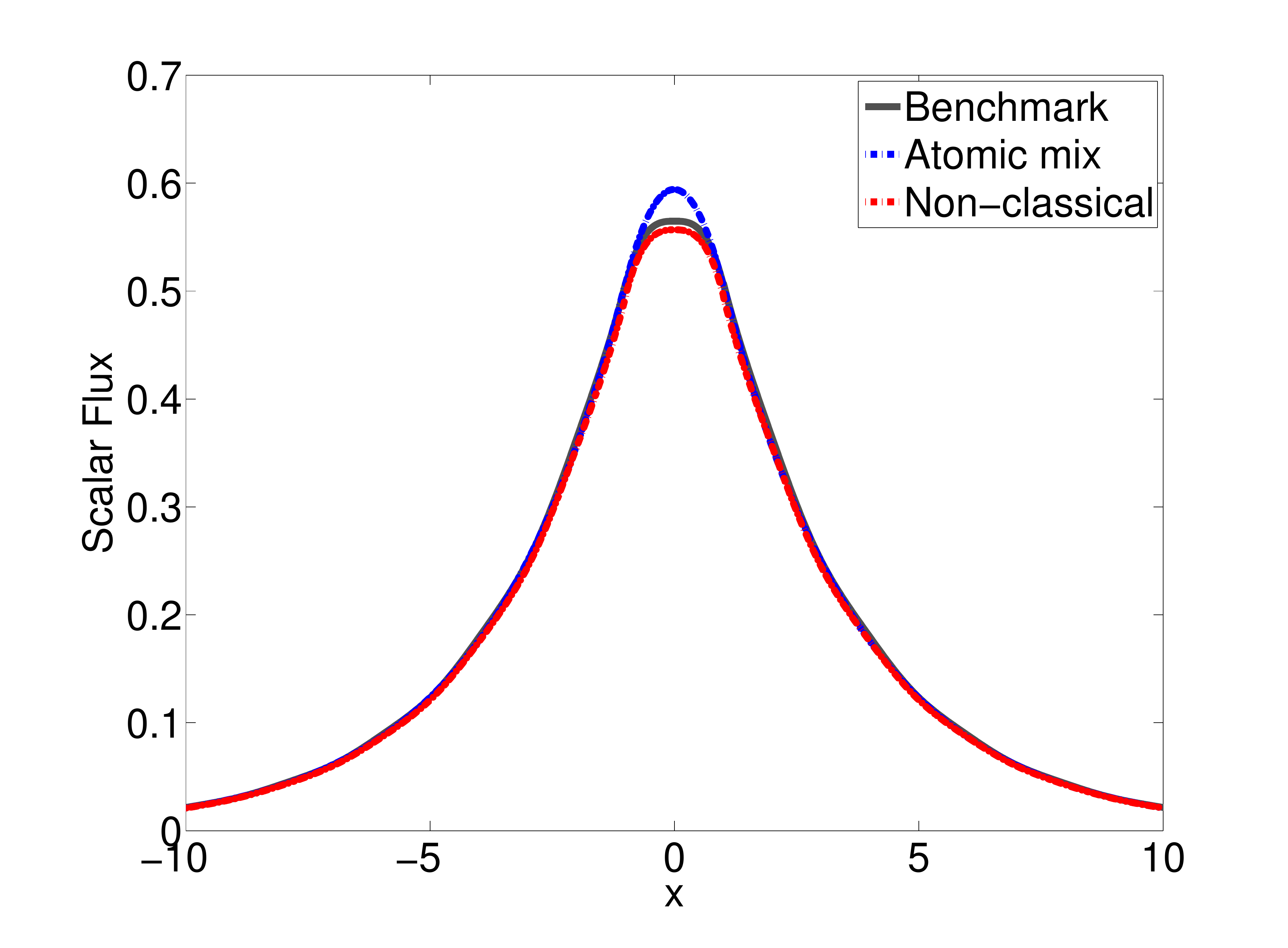}
        \caption{Problem set B, $c=0.5$}
        \label{fig:EP05b}
    \end{subfigure}        
    \\
    \begin{subfigure}[b]{0.495\textwidth}
        \centering
        \includegraphics[width=\textwidth]{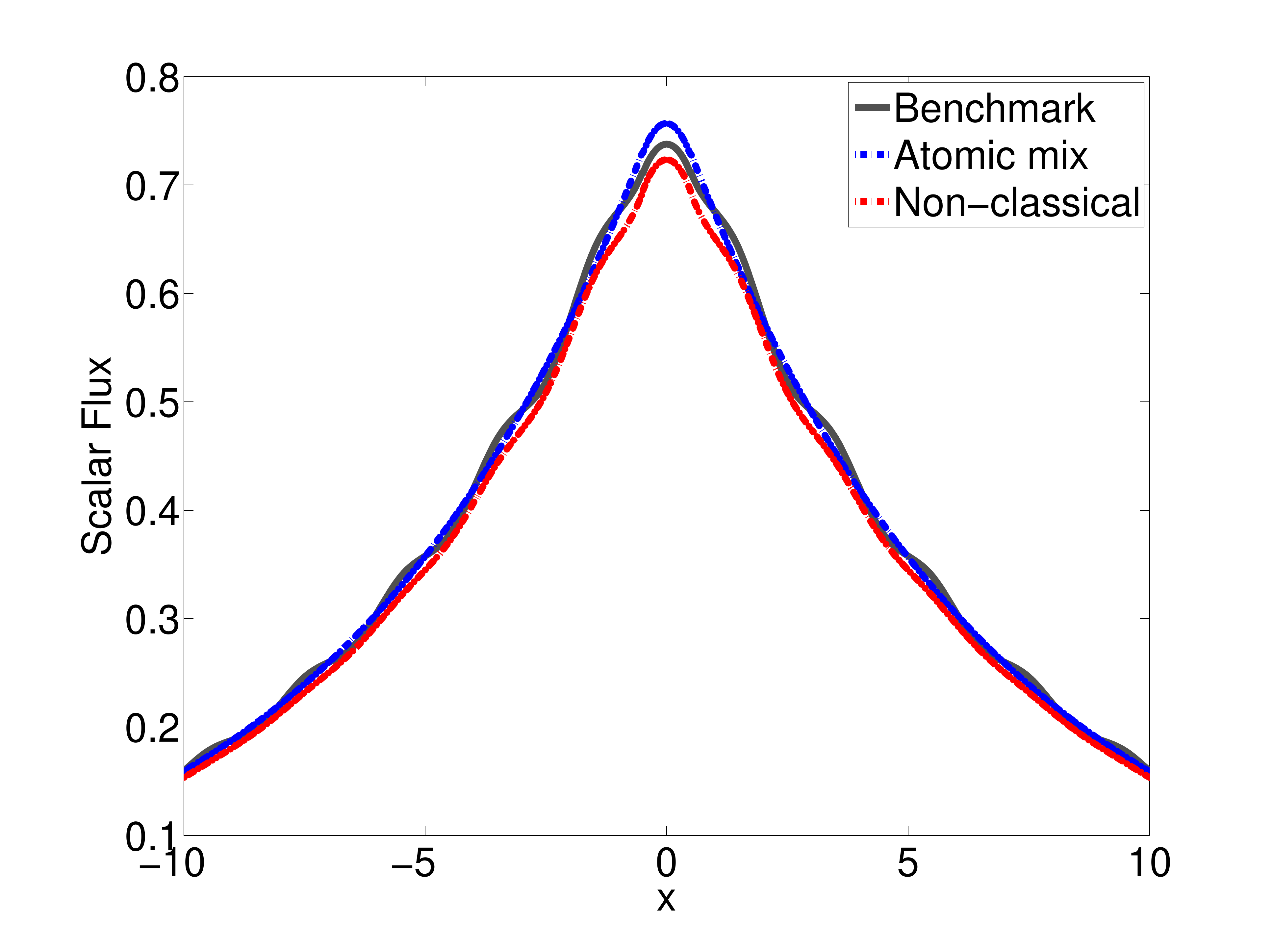}
        \caption{Problem set A, $c=0.9$}
        \label{fig:EP05a}
    \end{subfigure}
    \hfill
	 \begin{subfigure}[b]{0.495\textwidth}
        \centering
        \includegraphics[width=\textwidth]{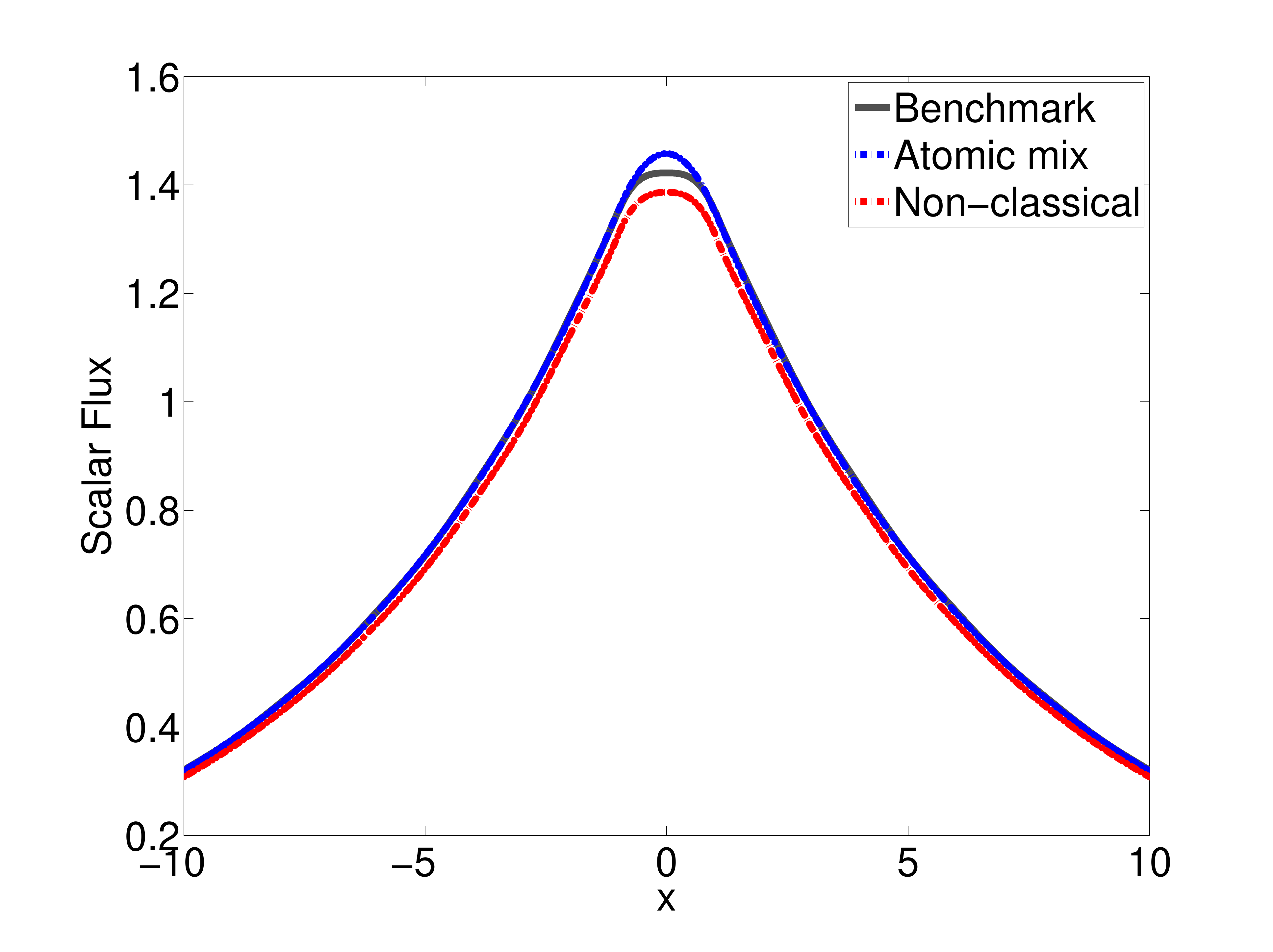}
        \caption{Problem set B, $c=0.9$}
        \label{fig:EP05b}
    \end{subfigure}        
    \caption{A comparison of the ensemble-averaged scalar fluxes of different models for problem sets \textbf{A} and \textbf{B}, with different scattering ratios $c$}
    \label{fig:Comp09}
\end{figure}

\begin{figure}
    \centering
    \begin{subfigure}[b]{0.495\textwidth}
        \centering
        \includegraphics[width=\textwidth]{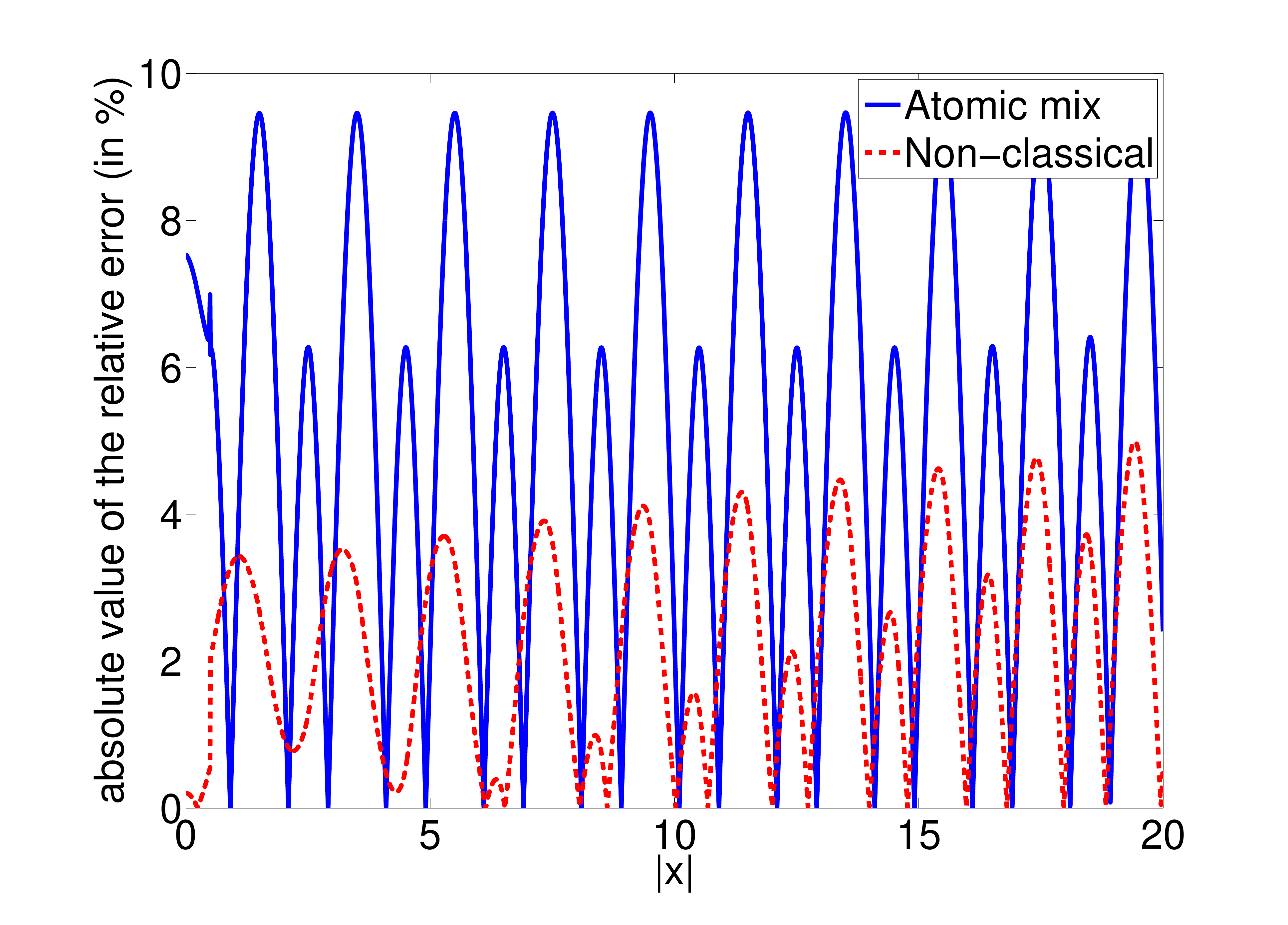}
        \caption{Problem set A, $c=0.1$}
        \label{fig:EP01a}
    \end{subfigure}
    \hfill
    \begin{subfigure}[b]{0.495\textwidth}
        \centering
        \includegraphics[width=\textwidth]{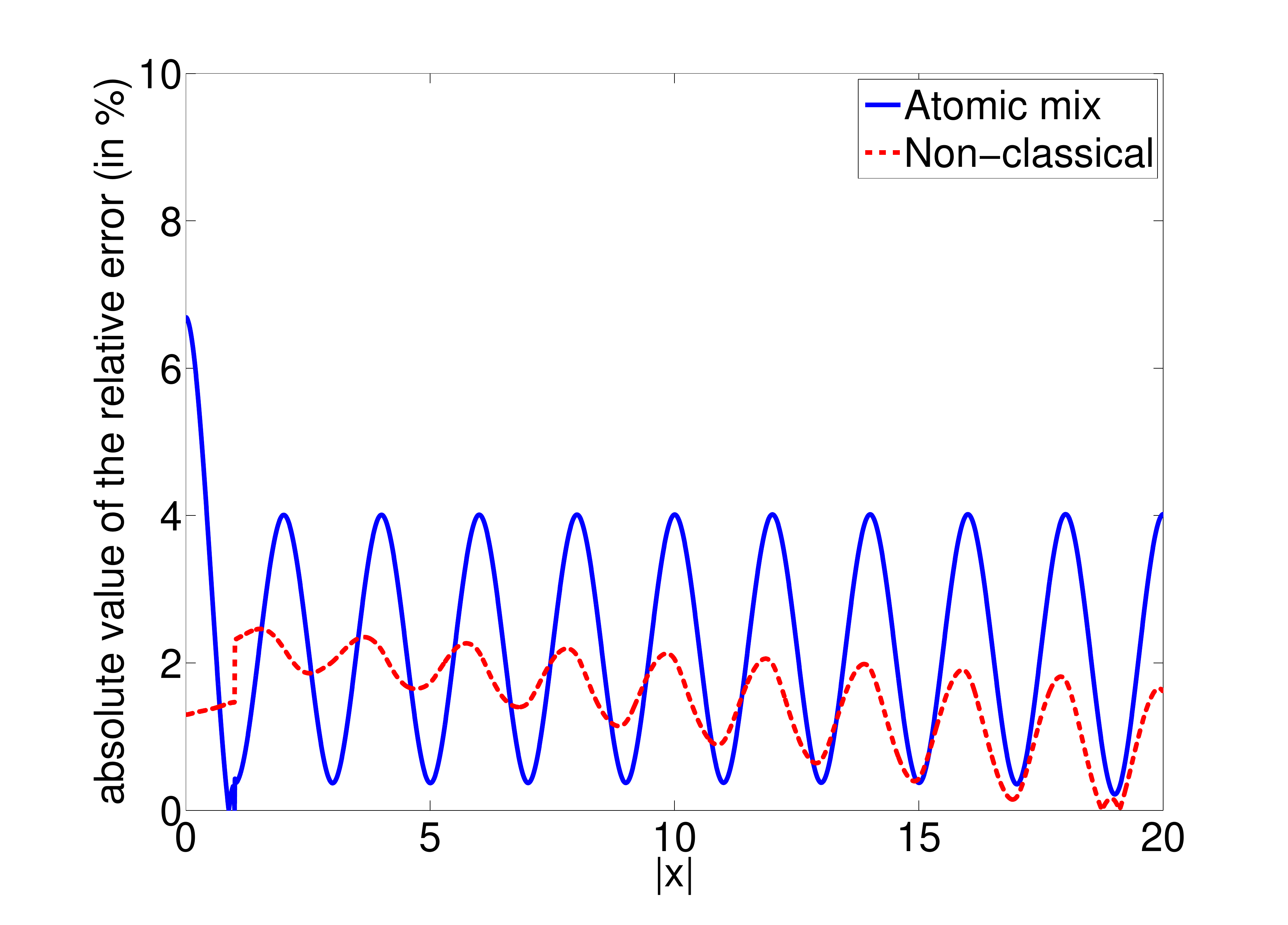}
        \caption{Problem set B, $c=0.1$}
        \label{fig:EP01b}
    \end{subfigure}
    \\
    \begin{subfigure}[b]{0.495\textwidth}
        \centering
        \includegraphics[width=\textwidth]{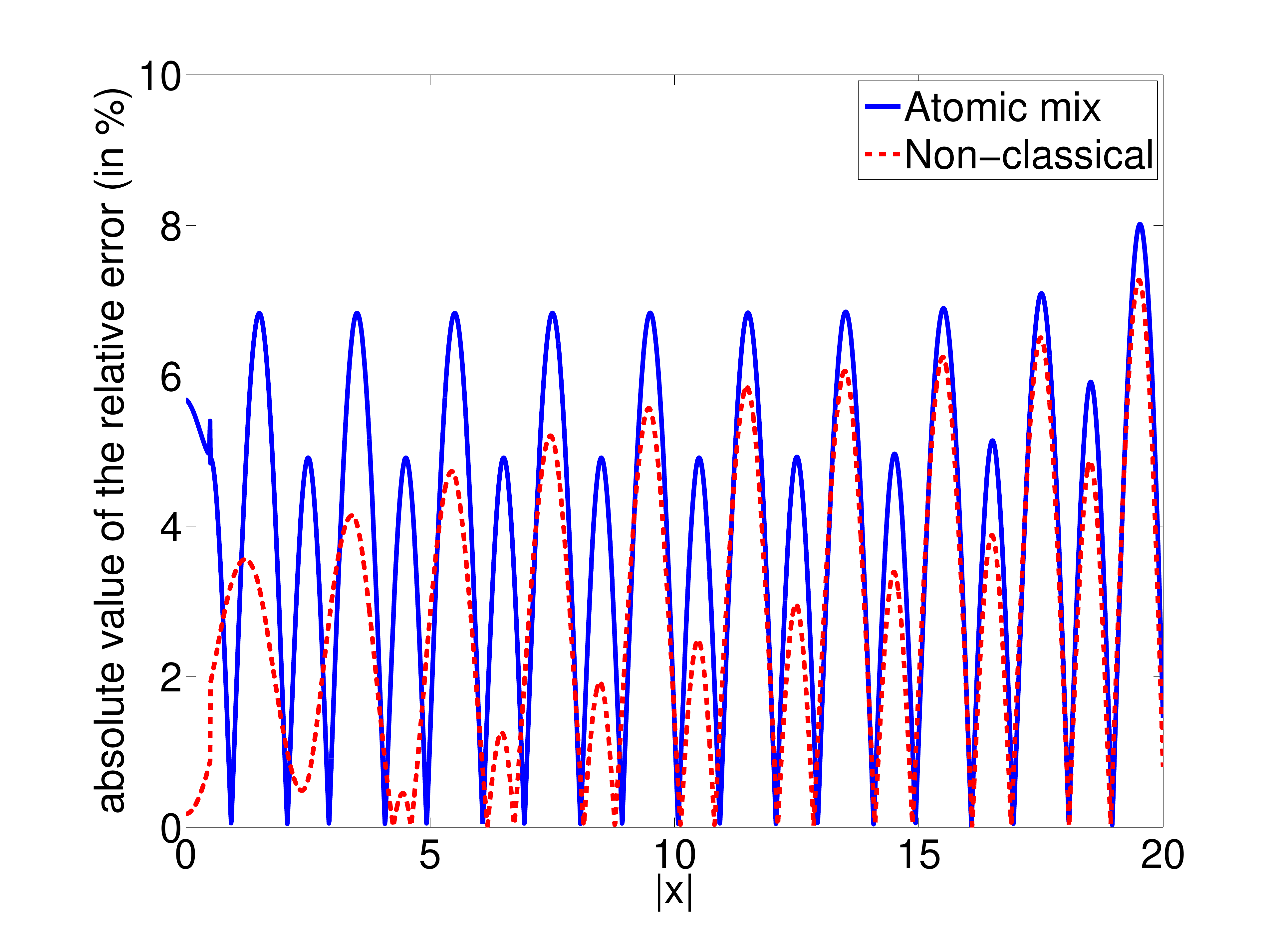}
        \caption{Problem set A, $c=0.5$}
        \label{fig:EP05a}
    \end{subfigure}
    \hfill
	 \begin{subfigure}[b]{0.495\textwidth}
        \centering
        \includegraphics[width=\textwidth]{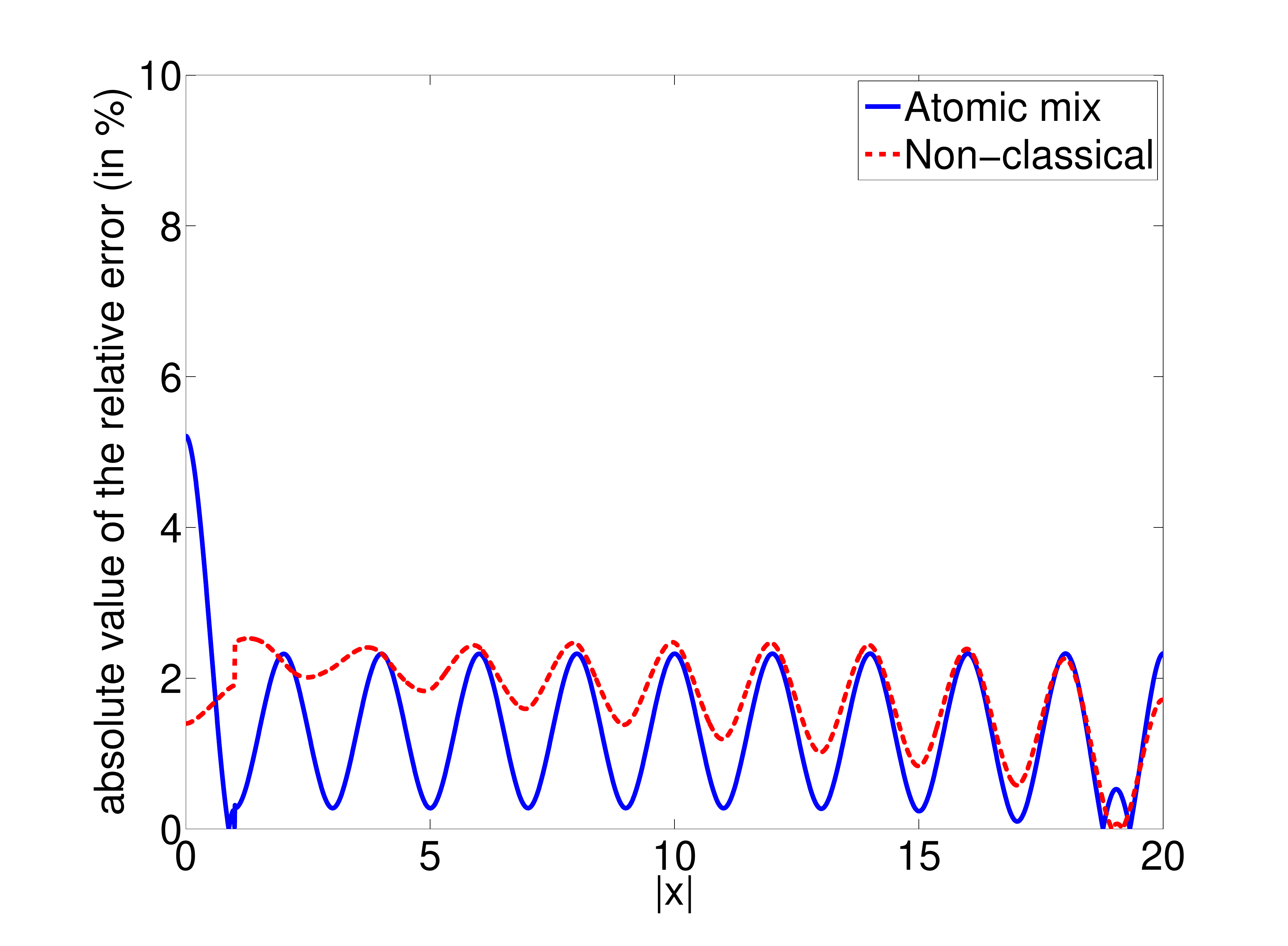}
        \caption{Problem set B, $c=0.5$}
        \label{fig:EP05b}
    \end{subfigure}        
    \\
    \begin{subfigure}[b]{0.495\textwidth}
        \centering
        \includegraphics[width=\textwidth]{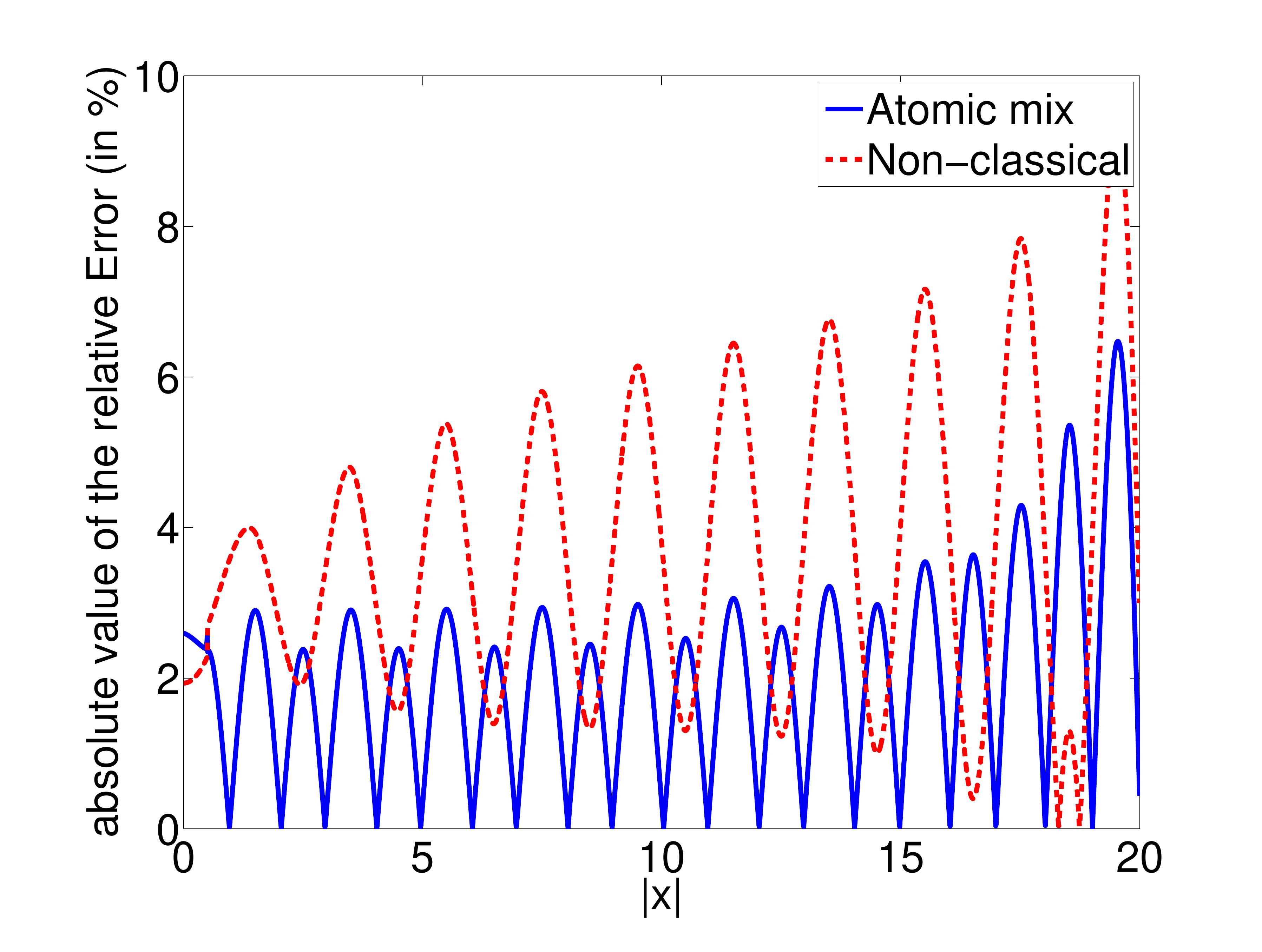}
        \caption{Problem set A, $c=0.9$}
        \label{fig:EP05a}
    \end{subfigure}
    \hfill
	 \begin{subfigure}[b]{0.495\textwidth}
        \centering
        \includegraphics[width=\textwidth]{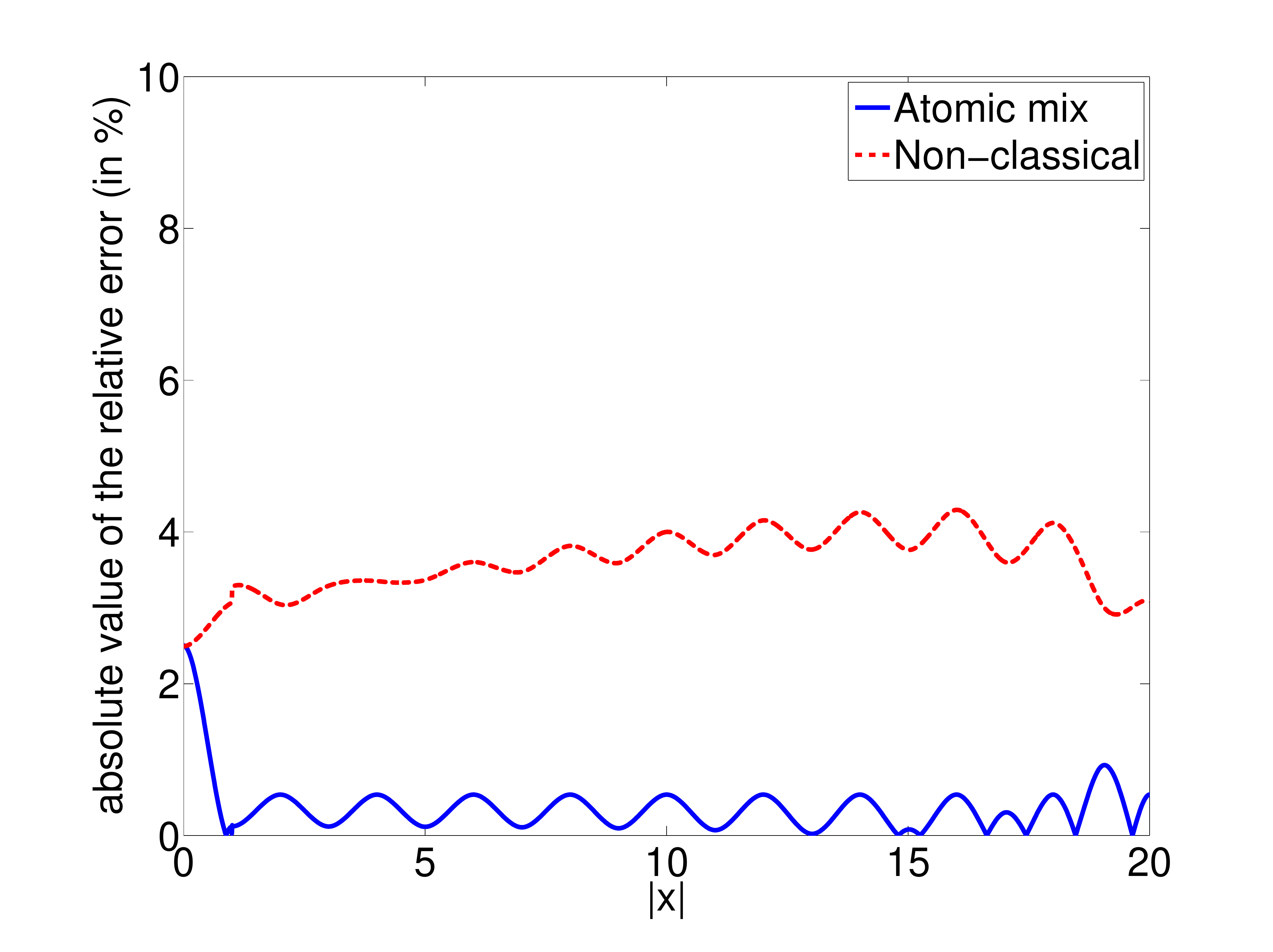}
        \caption{Problem set B, $c=0.9$}
        \label{fig:EP05b}
    \end{subfigure}        
    \caption{Absolute values of the atomic mix and non-classical relative errors with respect to the benchmark solutions for problem sets \textbf{A} and \textbf{B}}
    \label{fig:ErrorPlots}
\end{figure}

Figure \ref{fig:Comp09} presents the ensemble-averaged scalar fluxes obtained with each model for problem sets \textbf{A} and \textbf{B}. For a better analysis of these results, we define the relative errors of the models with respect to the benchmark solutions as
\begin{subequations}
\begin{align}
 Error_{AM}&= \frac{\bl\phi_{AM}\bg-\bl\phi_B\bg}{\bl\phi_B\bg}=\text{Atomic Mix Relative Error},\\
 Error_{NC}&= \frac{\bl\phi_{NC}\bg-\bl\phi_B\bg}{\bl\phi_B\bg}=\text{Non-Classical Relative Error}.
\end{align}
\end{subequations}
The absolute values of these relative errors (in $\%$) are plotted in Figure \ref{fig:ErrorPlots}.

The benchmark solutions for both sets of problems present a sinuous shape, due to the periodic structure of the random systems. This is significantly noticeable in problem set \textbf{A}, in which the amount of material 1 in the ``source region" $x\in[-0.5,0.5]$ varies in different realizations. In problem set \textbf{B}, all realizations have the same amount of material 1 in the source region $x\in[-1,1]$, leading to smoother benchmark solutions. Nevertheless, we can identify the sinuous behavior of the solutions in problem set \textbf{B} from the error plots in Figure \ref{fig:ErrorPlots}. It is important to notice that the non-classical model is able to capture this sinuous behavior, while the atomic mix model yields a smooth curve.

The error plots in Figure \ref{fig:ErrorPlots} indicate that, for 5 of the 6 test problems, the non-classical model outperforms the atomic mix model in estimating the flux around the origin. Table \ref{tab1} confirms this result, showing that atomic mix systematically overestimates the scalar flux at $x=0$. The data also shows that the non-classical model presents better accuracy at the origin, although it tends to slightly underestimate the maximum values. As the solution moves away from the center of the system, the non-classical model outperforms the atomic mix model for scattering ratio $c=0.1$, while atomic mix performs better in the case of $c=0.9$. 
\begin{table}[tbh]
\centering
\caption{\bf Ensemble-averaged scalar fluxes and relative errors at $x=0$}
\label{tab1} 
\vspace{14pt}
\begin{tabular}{||c|c||c|c|c||c|c||} \hline \hline
\textbf{Set}  & $c$ & $\bl\phi_B\bg$ & $\bl\phi_{AM}\bg$ &$\bl\phi_{NC}\bg$ & $Error_{AM}~(in~\%)$ & $ Error_{NC}~(in~\%)$\\ \hline\hline
& 0.1 & 0.2167 & 0.2331 & 0.2172 & 7.5390  & 0.2022 \\
 \cline{2-7}  
\textbf{A} & 0.5 & 0.3051 & 0.3224 & 0.3045 & 5.6902 & -0.1775 \\
 \cline{2-7}  
& 0.9 & 0.7377 & 0.7568 & 0.7234 & 2.6001 & -1.9339 \\
 \hline  
& 0.1 & 0.3922 & 0.4184 & 0.3871 & 6.6961 & -1.2983 \\
 \cline{2-7}  
\textbf{B} & 0.5 & 0.5648 & 0.5942 & 0.5569 & 5.2164 & -1.3978 \\
 \cline{2-7}  
& 0.9 & 1.4222 & 1.4578 & 1.3868 & 2.5048 & -2.4937 \\
 \hline\hline  
  \end{tabular}
\end{table}

Finally, we can see from Figure \ref{fig:ErrorPlots} that the accuracy of atomic mix increases as $c$ increases. The opposite happens with the non-classical model: its accuracy decreases as the scattering ratio increases. This effect might be caused, in part, by the numerical diffusion term in our HLL finite volume scheme, which smooths the solution. Further investigation in this matter shall be performed.

\section{Conclusion}\label{sec4}
\setcounter{section}{4}
\setcounter{equation}{0} 

This paper presents a first investigation of the accuracy of the non-classical transport theory in estimating the ensemble-averaged scalar flux in 1-D random media. In this preliminary work, simplifying assumptions were made: (i) the 1-D system is periodic, with randomness arising from randomly placing the arrangement in the infinite line; and (ii) transport takes place in rod geometry. To the best of our knowledge, this is the first time numerical solutions for the non-classical {\em transport} equation have been provided.

We show that, for the test problems presented here, the non-classical theory generally represents a more accurate alternative to the atomic mix model. It provides better estimates of the
maximum values of the scalar flux around the origin, and qualitatively preserves the sinuous shape of the solution. Moreover,
as the solution moves away from the center of the system, the non-classical model outperforms atomic mix for problems with low scattering.

This gain in accuracy comes at a cost: the path-length distribution function $p(s)$ [and its corresponding $\Sigma_t(s)$] must be known in order to solve the non-classical transport equation. However, considering that non-classical transport takes place in important nuclear applications (such as in Pebble Bed Reactor cores), it is our expectation that the gain in accuracy will prove the extra work worthwhile. In particular, the non-classical theory represents an alternative to current methods that might yield more accurate estimates of the eigenvalue and eigenfunction in a criticality calculation. 

Immediate future work includes: (i) dropping the rod geometry assumption, to generate slab geometry results; (ii) generalizing the parameters of the periodic random media, to obtain a more complete set of solutions; (iii) dropping the periodic assumption, to investigate results in more realistic random media. We point out that step (iii) cannot be performed with the analytical approach to obtain $p(s)$ presented in this paper. It requires either a numerical approach to estimate $p(s)$, or a (much) more complex mathematical theory.

\section*{Acknowledgments}
Richard Vasques would like to thank CNPq - Conselho Nacional de Desenvolvimento Cient\' ifico e Tecnol\' ogico for the financial support. The work of Kai Krycki was funded by the Excellence 
Initiative of the German federal and state governments.


\pagebreak

\setcounter{footnote}{0}
\begin{center}
{\bf CORRIGENDUM}\\
{\small January 30, 2016}\vspace{10pt}\\

{\bf Richard Vasques $^{\dagger,}$}\footnote{Email: \texttt{richard.vasques@fulbrightmail.org}} , {\bf Kai Krycki $^\ddagger$}\vspace{10pt}\\

\em {\bf $^\dagger$}University of California, Berkeley\\
 Department of Nuclear Engineering\\
 4155 Etcheverry Hall, Berkeley, CA 94720-1730\vspace{10pt}\\

 {\bf $^\ddagger$}Aachen Institute for Nuclear Training GmbH \\
 Jesuitenstraße 4, 52062 Aachen, Germany \em
\end{center}

Due to an oversight, an incorrect equation in this paper was printed as Eq.\ (2.7). The correct Eq.\ (2.7) is given below:

\setcounter{section}{2}
\setcounter{equation}{6}
\begin{align}
p(s)&=  \left\{
\begin{array}{ll}
\frac{\Sigma_{t1}}{\ell}[(2n+1)\ell-s]e^{-\Sigma_{t1}(s-n\ell)}, & \text{if } 2n\ell\leq s\leq (2n+1)\ell
\\ & \\
\frac{\Sigma_{t1}}{\ell}[s-(2n+1)\ell]e^{-\Sigma_{t1}(s-(n+1)\ell)}, &\text{if } (2n+1)\ell\leq s\leq 2(n+1)\ell\\
\end{array} 
\right. 
\end{align}
The numerical results presented in the figures and tables of this paper were obtained with the correct Eq.\ (2.7).

The same incorrect equation was printed in the published version of this paper (on CD-ROM) in the {\em Proceedings of the Joint International Conference on Mathematics and Computation, Supercomputing in Nuclear Applications and the Monte Carlo Method, M\&C 2015}, held in Nashville, TN, April 19-23 (2015). The corrigendum can be found in the website of the author Richard Vasques: \href{http://ricvasques.github.io}{https://ricvasques.github.io/}. 

\setlength{\baselineskip}{12pt}





\end{document}